# Control of Arrival Time using Structured Wave Packets

T. A. Saxton[+] and A. L. Harris[*]

Physics Department, Illinois State University, Normal, IL, USA 61790

## Abstract

Scattering dynamics are examined for Gaussian and non-Gaussian wave packets with identical momentum densities. Average arrival time delays, dwell times, and phase time delays are calculated for wave packets scattering from a square barrier, and it is shown that the non-Gaussian wave packets exhibit different average arrival time delays than the Gaussian wave packets. These differences result from the non-linear terms in the momentum wave function phase of the non-Gaussian wave packets, which alters the self-interaction times of the wave packets. Control of the average arrival time delay can be achieved through adjustment of the momentum wave function phase, independent of wave packet energy and momentum density.

## 1. Introduction

What features of a quantum mechanical wave packet influence its interaction time with a localized potential and can a wave packet's arrival time at a given location be controlled? The answers to these questions have both fundamental importance, as well as applications in fields such as microscopy [1–3], attosecond science [4–6], materials science [7–9], and electronics [10,11]. The idea of interaction time relates closely to the concepts of traversal time and tunneling time, in which one is interested in the duration of time a particle spends in the interaction region and/or its arrival time at a given spatial location. For potential barriers, arrival of the particle beyond the interaction region may be a result of tunneling through the potential barrier or transmission over the potential barrier. Unfortunately, despite decades of investigations, there is no unique definition for a tunneling or arrival time because time is a parameter, not an observable, in quantum mechanics [12]. This has led to countless suggestions and definitions for calculating and measuring tunneling and arrival times [4,12–24].

*corresponding author alharri@ilstu.edu
+ Current Address Department of Physics, University of Notre Dame, Notre Dame, IN 46556

For tunneling time, even the basic question of whether the tunneling process occurs instantaneously or over a finite time interval has not yet been answered [4–6,15,25–28]. The question has been explored in the context of attosecond tunneling spectroscopy and attoclock experiments [4,6,26,29], as well as condensed matter applications [30,31]. Some experiments have yielded tunneling times of zero [4,29,32], while others resulted in non-zero tunneling times [6,26]. Predictions of recent theoretical models are also mixed, with models such as the time-dependent Schrödinger equation and numerical attoclock simulations showing zero tunneling time [5,27,32–36], while earlier models predict non-zero tunneling times [12,18,37,38].

The interpretation of arrival times has also received its share of attention, especially with regards to what is known as the Hartman Effect. In 1962, Hartman showed that the transmission time for a tunneling Gaussian particle becomes independent of barrier width for wide barriers [38]. While this might seem to imply the possibility of superluminal velocity, there is no violation of causality and several explanations of the transmission time saturation have been proposed. These include wave packet reshaping, [38–42], saturation of stored energy (photons) or integrated probability density (electrons) within the barrier [15,43,44], and interference effects [45–48].

In recent years, renewed interest in tunneling and arrival time has been spurred by technological developments and applications in areas such as electronics, imaging, and quantum information. Electronic devices such as Josephson junctions, tunnel diodes, and other nanoelectronic devices have electron tunneling at their core [2,8,9], where it is directly related to their functional speed. Further miniaturization of these devises requires a thorough understanding of tunneling dynamics [9]. In imaging applications, electron microscopes are used to obtain detailed structural information about biological, molecular, and nanostructure samples. Enhancements in microscope resolution require tunneling techniques, as do new methods that utilize electron wave function phase [2]. In the quantum information and quantum computing realm, the tunneling of Bose-Einstein condensates through optical lattices is being explored [8], as is tunneling between quantum dots for use in computation and information storage [49]. Initial experiments show the possibility of controlling tunneling time in this environment [8].

In parallel with the recent interest in tunneling and arrival times has been the experimental generation of electron wave packets with non-Gaussian spatial profiles [1,50], such as electron Airy beams [1] and electron vortex beams [50,51]. These new spatially structured electron wave packets have unique properties such as quantized orbital angular momentum, self-acceleration, self-healing, and minimal dispersion [1,50,52]. Their proposed uses include the study of fundamental atomic properties [53–55], control and rotation of nanoparticles [53,54,56,57], and electron microscopy [53,58,59]. For example, super resolution and light sheet optical microscopy have become powerful tools that rely on the properties of spatially structured light beams to achieve resolution beyond the diffraction limit [60–62]. It is possible that spatially structured electron beams could lead to similar improvements in electron microscopy [2].

The experimental realization of spatially structured electron beams, combined with the fundamental importance and numerous applications of tunneling and arrival times, leads to the unique opportunity to combine these areas of study to directly examine how individual wave function properties affect interaction times. In particular, the comparison of scattering dynamics for carefully chosen Airy and Gaussian wave forms can provide insight into the question of how wave function phase, spatial density, and momentum density influence arrival time, which in turn provides a test of the various arrival time definitions. Additionally, the investigation of non-Gaussian wave packet tunneling and transmission dynamics is useful for the proposed applications that rely on the use and control of these recently realized wave packets.

To date, most studies of tunneling and arrival time have been performed using Gaussian wave packets. In this case, any change to the spatial density of the wave packet also changes its momentum density, making direct study of how wave packet properties influence arrival time difficult. An Airy wave packet, on the other hand, has a Gaussian momentum density, but a non-Gaussian spatial density and can therefore be designed with a momentum density identical to that of a spatial Gaussian wave packet. This allows for direct comparison of the effect of spatial and momentum density on arrival time. In addition, Airy and Gaussian wave packets differ in momentum wave function phase, with the Gaussian wave function phase being linear, while the Airy wave function phase is cubic. Comparison of carefully designed Airy and Gaussian wave packets whose primary difference is momentum phase allows for direct study of the role of phase in arrival time.

Here, we use the average arrival time delay [13], phase time delay [18–20], and dwell time [14] to quantify the interaction time for both over-the-barrier scattering and tunneling. We show that the average arrival time delays and phase time delays of wave packets with identical momentum densities are *not* necessarily identical, and we determine under what conditions the arrival time of an Airy wave packet differs from that of a Gaussian wave packet with identical momentum density. We also show that the arrival time of a general non-Gaussian wave packet can be controlled through adjustment of the momentum wave function phase, independent of the incident wave packet's energy or momentum density.

## 2. Theory

Consider the one-dimensional scattering of a wave packet with mean incident momentum $\hbar k_0$ from a square barrier of height $V_0$ and width $L$, as shown in Figure 1. During the collision, the wave packet undergoes reflection, transmission, and tunneling. Classically, if the particle's incident energy $E$ is less than the barrier height, it can only be reflected. For classical transmission beyond the barrier region to occur, the particle's energy needs to be greater than the barrier height. However, for quantum mechanical particles, transmission can occur even when the particle's mean energy is less than the barrier height, resulting in a non-zero probability of finding the particle beyond the barrier region. Conversely, quantum mechanical wave packets with mean energy above the barrier height will experience some reflection from the barrier, resulting in transmission probability less than 1.

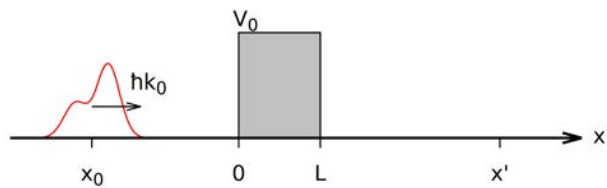

Figure 1 Wave packet with mean momentum $\hbar k_0$ and mean position $x_0$ incident on a square barrier potential with width $L$ and height $V_0$. Average arrival time is calculated at $x'$.

Reflection and transmission coefficients can be used to quantify the probability of finding the particle to the right or left of the barrier long after the scattering has occurred, and these asymptotic quantities provide information about the spatial and momentum features of the interaction. However, they do not provide information about the time of interaction. To gain insight into the wave packets' temporal interactions with the barrier, it is necessary to examine quantities that characterize the duration of the scattering process, which requires the calculation of time-dependent wave functions.

We calculate time-dependent spatial wave functions with our Path Integral Quantum Trajectory (PIQTr) model [63] in which an initial state wave function is propagated in time by iterating the equation

$$\psi(x_b, t_b) = \int_{-\infty}^{\infty} K(x_a, x_b, t_a, t_b)\psi(x_a, t_a)dx_a, \qquad (1)$$

for small time steps. The initial state wave function at position $x_a$ and time $t_a$ is $\psi(x_a, t_a)$. The propagator $K(x_a, x_b, t_a, t_b)$ is written in terms of the classical action, and the result is the time-evolved wave function $\psi(x_b, t_b)$ at position $x_b$ and time $t_b$. To directly determine how wave packet properties, such as spatial density, wave function phase, and spatial and momentum uncertainty affect interaction time, we compare three specific wave packets: the Gaussian, Airy, and inverted Airy wave packets.

## 2.1 Gaussian Wave Packets

The Gaussian wave packet is the most common localized wave packet used for the study of scattering and its spatial wave function is given by

$$\psi^G(x, 0) = \frac{e^{-(x-x_0)^2/2\sigma^2}}{(\pi\sigma^2)^{1/4}} e^{ik_0(x-x_0)}, \qquad (2)$$

with standard deviation $\sigma$ and initial mean position $x_0$. It is the minimum uncertainty wave packet because it satisfies the lower bound of the uncertainty relation

$$\Delta x \Delta p \geq \hbar/2. \qquad (3)$$

For the Gaussian wave packet, the equality in Eq. (3) holds, and there is a one-to-one relationship between the spatial and momentum uncertainty. A larger spatial uncertainty always results in a smaller momentum uncertainty and vice versa. The corresponding Gaussian momentum wave function is itself a Gaussian

$$\varphi^G(p,0) = \frac{\sigma^{1/2}}{\pi^{1/4}} e^{-ipx_0} e^{-\sigma^2(p-k_0)^2/2}, \tag{4}$$

with linear phase

$$\phi^G = -px_0. \tag{5}$$

**2.2 Airy Wave Packets**

Recently, non-Gaussian localized wave packets have been experimentally generated using electrons [1,51,64,65]. One such wave packet is the truncated Airy wave function [1,66]

$$\psi^A(x,0) = (8\pi\alpha)^{1/4} e^{-\alpha^3/3} Ai(x-x_0) e^{\alpha(x-x_0)} e^{ik_0(x-x_0)}, \tag{6}$$

with truncation parameter $\alpha$ and Airy function $Ai(x-x_0)$. The Airy wave function without the truncation term is a solution to the free particle Schrödinger equation [52], but like its more commonly used plane wave counterpart, has infinite transverse extent and infinite energy. The truncation term in Eq. (6) yields a finite width wave packet with finite energy that is square normalizable.

Airy wave functions have many unique properties, including self-acceleration, self-healing, and minimal spreading [1,52,66] and were first described by Berry and Balazs [52]. They showed that the ideal Airy wave function is non-dispersive and that its individual lobes move along a parabolic trajectories. However, the expectation value of the Airy's position does not exhibit acceleration and therefore the wave packet's mean position follows the same linear path as any other free particle. These unique properties generally persist for the truncated Airy wave packet, as well [1,66]. Self-acceleration of the individual lobes is observed, as is minimal dispersion and self-healing. While these features are intriguing by themselves, we are primarily interested in the truncated Airy wave function's Gaussian momentum density because it provides a straightforward means to directly study the role of position and momentum uncertainty, spatial and momentum density, and momentum wave function phase in scattering dynamics through comparison of Gaussian and Airy wave packets.

The Airy momentum wave function is given by

$$\varphi^A(p,0) = \frac{(8\pi\alpha)^{1/4}}{\sqrt{2\pi}} e^{-\alpha(p-k_0)^2} e^{-ipx_0} e^{i\left[\frac{(p-k_0)^3}{3} - \alpha^2(p-k_0)\right]}, \tag{7}$$

with cubic phase

$$\phi^A = \phi^G + \left[\frac{(p-k_0)^3}{3} - \alpha^2(p-k_0)\right]. \tag{8}$$

The modulus squared of Eq. (7) is clearly seen to be Gaussian, allowing for the standard deviation of the Gaussian wave packet and the truncation parameter of the Airy wave packet to be chosen in such a way that their initial momentum densities $|\varphi(p,0)|^2$ and momentum uncertainties $\Delta p$ will be identical ($\alpha = \frac{\sigma^2}{2}$). However, their spatial densities $|\psi(x,0)|^2$, position uncertainties $\Delta x$, and momentum wave function phases $\phi^{A,G}$ will be different. For example, an Airy wave packet with $\alpha = 0.317$ has the same momentum density and momentum uncertainty as a Gaussian wave packet with $\sigma = 0.795$, but a different spatial density, position uncertainty, and momentum wave function phase (see Table 1 and Figure 3).

Unlike the Gaussian wave packet, the Airy wave packet is not a minimum uncertainty wave packet and there is not a one-to-one relationship between position and momentum uncertainty. For most values of $\alpha$, a given position uncertainty leads to two different momentum uncertainties, such that two Airy wave functions can be designed with identical position uncertainties $\Delta x$, but different momentum uncertainties $\Delta p$. For example, Airy wave packets with $\alpha = 0.317$ and $\alpha = 1.5$ have identical position uncertainties but different momentum uncertainties, spatial densities, and momentum wave function phases. Figure 2 shows the spatial and momentum uncertainties of the Gaussian and Airy wave packets as a function of the standard deviation and truncation parameter. For large values of $\alpha$ and $\sigma$, the position and momentum uncertainties of the Airy and Gaussian wave packets become identical, however their momentum wave function phases and spatial densities remain different.

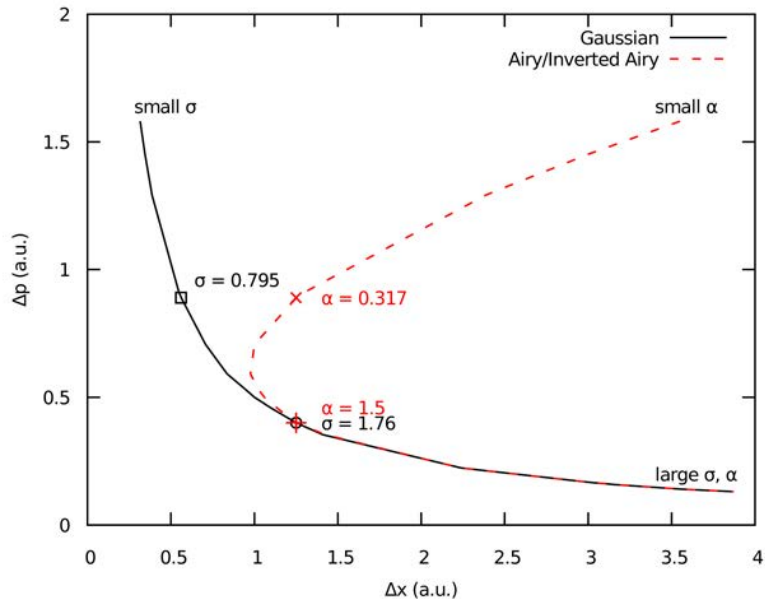

Figure 2 Position and momentum uncertainties for the Gaussian, Airy, and inverted Airy wave packets as a function of standard deviation $\sigma$ and truncation parameter $\alpha$.

### 2.3 Inverted Airy Wave Packets

Unlike the Gaussian wave function, the Airy wave packet's spatial density is not symmetric (see Figure 3) and the spatial orientation of the wave packet relative to its momentum direction is physically important. An Airy wave packet with its largest peak to the right and momentum to the right behaves differently than an Airy wave packet with its largest peak to the left and momentum to the right. The latter case is a spatial reflection of the Airy wave packet and is referred to as the inverted Airy wave packet.

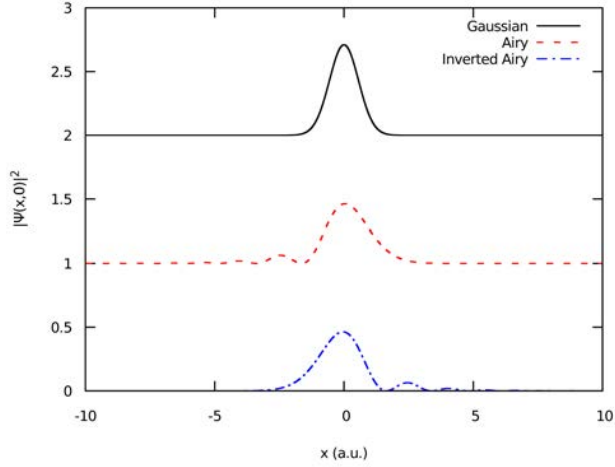

Figure 3 Spatial densities for the Gaussian ($\sigma = 0.795$), Airy ($\alpha = 0.317$), and inverted Airy ($\alpha = 0.317$) wave packets with $x_0 = 0$ a.u. The Gaussian and Airy curves have been shifted above the x-axis for clarity.

The inverted Airy momentum wave function is identical to that of the Airy wave function, except for a change in sign in the non-Gaussian term of the phase

$$\varphi^{IA}(p,0) = \frac{(8\pi\alpha)^{1/4}}{\sqrt{2\pi}} e^{-\alpha(p-k_0)^2} e^{-ipx_0} e^{-i\left[\frac{(p-k_0)^3}{3} - \alpha^2(p-k_0)\right]}, \tag{9}$$

with

$$\phi^{IA} = \phi^G - \left[\frac{(p-k_0)^3}{3} - \alpha^2(p-k_0)\right]. \tag{10}$$

Airy and inverted Airy wave functions with identical truncation parameters will always have the same spatial uncertainties, momentum uncertainties, and momentum densities, but different momentum wave function phases and spatial densities. Comparison of Airy and inverted Airy wave packet scattering can then be used to isolate the effects of momentum wave function phase and spatial density.

**2.4 Interaction Time**

Unfortunately, even the simple concept of when a particle arrives at a given spatial location still has no universally accepted definition within quantum mechanics [4,12–23]. While such an arrival time definition is straightforward for classical particles, quantum particle characteristics such as tunneling, interference, uncertainty, and dispersion complicate attempts to define a quantum mechanical arrival time. Many candidates are offered in the extensive literature that discusses the feasibility and applicability of such definitions [4,12–24], and we refer the interested reader to these more detailed analyses and references therein. Here, we have chosen three generally accepted quantities to characterize the interaction time. They are the average arrival time delay [13], the average dwell time of the particle in the barrier region [14], and the group (or phase time) delay [18–20].

The average arrival time $T(x')$ is a local quantity that averages the spatial density at a location $x'$ over time to yield an average time that the particle arrives at $x'$. It is given by [13]

$$T(x') = \int_0^\infty dt\, t P_{x'}(t), \tag{11}$$

where $P_{x'}(t)$ is the arrival time distribution

$$P_{x'}(t) = \frac{|\psi(x',t)|^2}{\int_0^\infty dt' |\psi(x',t')|^2}. \tag{12}$$

Since we are interested in the duration of the interaction, the average arrival time delay between interacting and non-interacting particles can be used to quantify the interaction time. The average arrival time delay is defined as

$$\Delta T^i = T^i_{int} - T^i_{free}, \tag{13}$$

where $T^i_{int}$ is the average arrival time of an interacting particle, $T^i_{free}$ is the average arrival time of an identical non-interacting free particle, and $i = G, A, IA$ refers to either a Gaussian, Airy, or inverted Airy wave packet, respectively. A positive value of $\Delta T^i$ indicates that the interacting particle was impeded by the barrier, while a negative value indicates that the particle was accelerated by its interaction with the barrier.

The group delay, or phase time delay, can be found through analytical analysis using the stationary phase approximation [18–20]. In this approximation, it is assumed that the phase of the momentum wave function is stationary at each barrier edge $x = 0$ and $x = L$, and a time of arrival at these two points is used to quantify the delay of the particle relative to an identical free particle. It is analogous to the average arrival time delay [13]. The phase time delays for the Airy $\Delta t_\zeta^A$ and inverted Airy $\Delta t_\zeta^{IA}$ wave packets have an additional term compared to the Gaussian phase time delay $\Delta t_\zeta^G$ such that their phase time delay either increases or decreases relative to the Gaussian wave packet (see Appendix A). They are given by

$$\Delta t_\zeta^A = \Delta t_\zeta^G + m \frac{(k_T - k_0)^2 - \alpha^2}{\hbar k_T} \tag{14}$$

for the Airy wave packet and

$$\Delta t_\zeta^{IA} = \Delta t_\zeta^G - m \frac{(k_T - k_0)^2 - \alpha^2}{\hbar k_T} \tag{15}$$

for the inverted Airy wave packet, where $\Delta t_\zeta^G$ is the Gaussian phase time delay time, $\hbar k_T$ is the average momentum of the transmitted wave packet, and $m$ is the mass of the particle. The additional term in the phase time delay results from the non-linear terms in these wave packets' momentum wave function phases. Physically, the phase time delay is a result of two independent contributions [15,67] – the dwell time $\tau_D$ and the self-interaction time $\tau_I$

$$\Delta t_\zeta = \tau_D + \tau_I. \tag{16}$$

The dwell time quantifies the total time a particle spends in the barrier region. It is a non-local quantity that averages the spatial density of the wave packet over the barrier region [14]

$$\tau_D = \int_{-\infty}^{\infty} dt \int_0^L dx |\psi(x,t)|^2. \tag{17}$$

The self-interaction time is a result of interference between the incident and reflected parts of the wave packets in the region to the left of the barrier and can be found by subtracting the dwell time from the phase time delay.

## 3. Results

### 3.1 Wide Barrier

We begin with the specific case of Gaussian, Airy, and inverted Airy wave packets scattering from a barrier of height $V_0 = 50$ a.u. and width $w = 1$ a.u. The wave packets are chosen such that they have identical momentum densities, but different spatial densities, position uncertainties, and momentum wave function phases. They are labeled as wave packets 1-3 in Table 1, which lists their position and momentum uncertainties and average arrival time delays. The initial mean position of the wave packets is $x_0 = -15$ a.u. and the initial mean momentum is $\hbar k_0 = 7.75$ a.u. The average arrival time delays are calculated at $x' = 1$ a.u., which is the right edge of the barrier, and $x' = 20$ a.u., which is sufficiently far from the barrier that the scattering process has been completed. The average arrival time delays at $x' = 1$ a.u. show that the Gaussian and inverted Airy wave packets are accelerated by their interaction with the barrier relative to an identical non-interacting free particle, which results in negative average arrival time delays. In contrast, the Airy wave packet is delayed by its interaction with the barrier. However, at the asymptotic location of $x' = 20$ a.u., all three wave packets arrive before their free particle counterparts. This is due to an increase in the mean momentum of the transmitted wave packet relative to the mean incident momentum, as discussed below.

Clearly, the average arrival time delays are different for wave packets 1-3, and we note that regardless of which position is used, the relative average arrival time delays between the wave packets are unchanged. The inverted Airy wave packet has the smallest average arrival time delay, indicating that its interaction time with the barrier is shortest and it arrives at $x'$ earliest. The Airy wave packet has the largest average arrival time delay and interaction time with the barrier and therefore arrives at $x'$ latest. The Gaussian wave packet's average arrival time delay lies in between the Airy and inverted Airy wave packets' average arrival time delays. Additional evidence of the variation in average arrival time delays can be observed in the time-dependent position and momentum densities, as well as the arrival time distributions shown in Figure 4.

|  | (1) Inverted Airy $\alpha = 0.317$ | (2) Gaussian $\sigma = 0.795$ | (3) Airy $\alpha = 0.317$ | (4) Inverted Airy $\alpha = 1.5$ | (5) Gaussian $\sigma = 1.76$ | (6) Airy $\alpha = 1.5$ |
|---|---|---|---|---|---|---|
| $\Delta x$ | 1.25 | 0.56 | 1.25 | 1.25 | 1.25 | 1.25 |
| $\Delta p$ | 0.89 | 0.89 | 0.89 | 0.4 | 0.4 | 0.4 |
| $\Delta T(x' = 1)$ | -0.8 | -0.3 | 0.3 | -0.2 | -0.2 | -0.2 |
| $\Delta T(x' = 20)$ | -1.4 | -0.9 | -0.3 | -0.3 | -0.3 | -0.3 |

Table 1 Position and momentum uncertainties and average arrival time delays for the wave packets discussed in the text and figures. Average arrival time delays are calculated for $x' = 1$ a.u., $x' = 20\ a.u.$, $x_0 = -15\ a.u.$, $V_0 = 50\ a.u.$, $\hbar k_0 = 7.75$ a.u., and $w = 1$ a.u.

The arrival time distributions $P_{x'}(t)$ (row 1) for wave packets 1-3 show signatures of the different average arrival times between the three wave packets, with the peaks of the inverted Airy wave packet distributions located at earlier times than the Gaussian and Airy wave packet distributions. This separation of the distributions leads to the quantifiable differences in average arrival time delays shown in Table 1. The shapes of the arrival time distributions are a result of the different spatial density profiles of the transmitted wave packets, which are altered from the incident wave packet densities due to the scattering process.

The time-dependent spatial densities for wave packets 1-3 are shown in row 2 of Figure 4, where the inverted Airy wave packet is observed to arrive at $x' = 1$ a.u. and $x' = 20$ a.u. before the Gaussian wave packet, while the Airy wave packet arrives after the Gaussian wave packet. Because Figure 4 and Table 1 show clear differences in average arrival time delays between wave packets with identical momentum densities, we conclude that a wave packet characteristic(s) other than momentum density determines the interaction time. The remaining differences between the wave packets are the spatial densities, position uncertainties, and momentum wave function phases, and a pair-wise comparison of wave packet scattering dynamics can be used to identify which of these characteristics causes the average arrival time delays to be different.

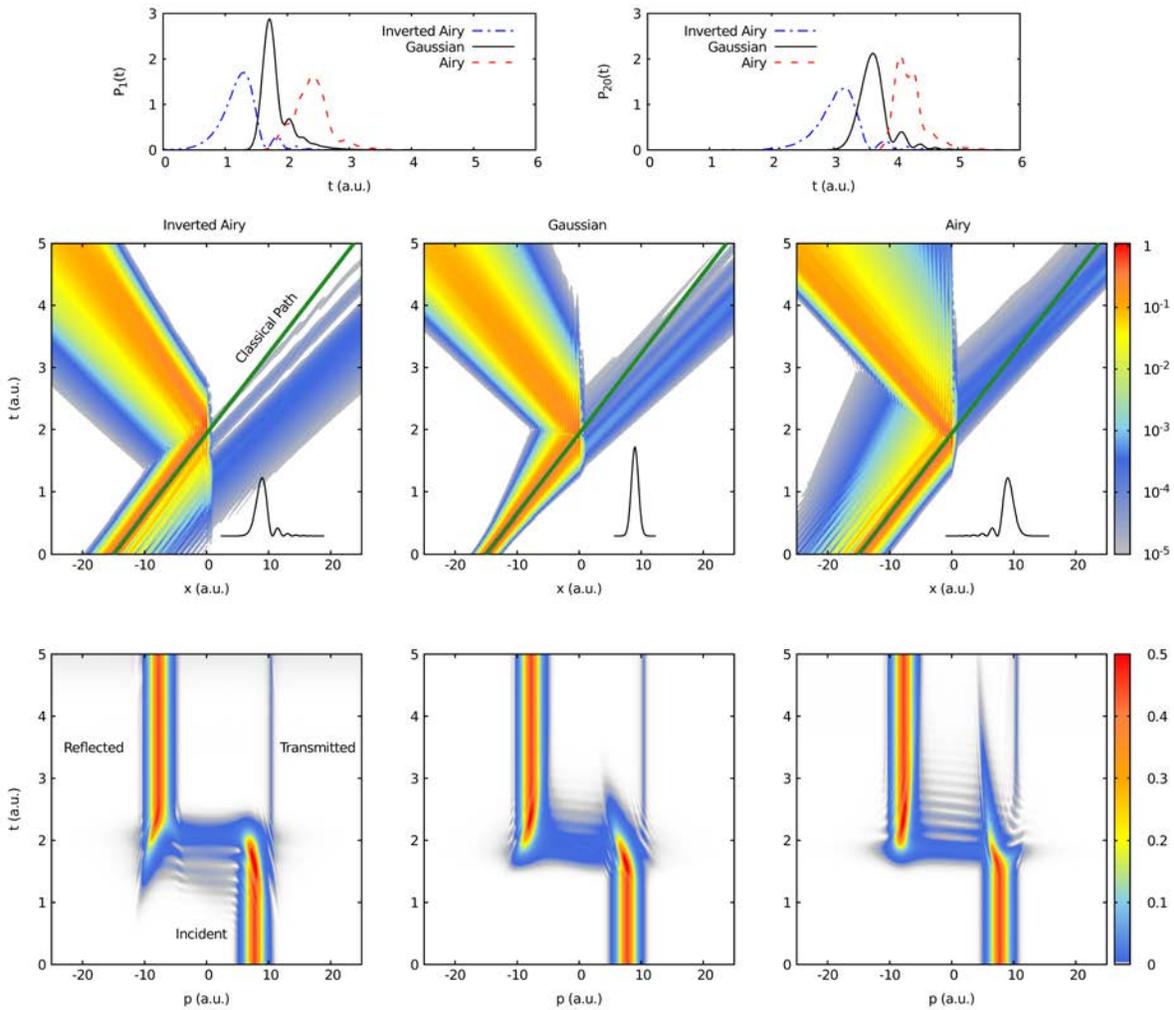

Figure 4 Row 1 – arrival time distributions at $x' = 1$ a.u. and $x' = 20\ a.u.$ for inverted Airy, Gaussian, and Airy wave packets interacting with a square barrier of height $V_0 = 50\ a.u.$ and width $w = 1\ a.u.$

The mean incident momentum is $\hbar k_0 = 7.75$ a.u. and initial mean position is $x_0 = -15\ a.u.$ Rows 2 and 3 – time-dependent position and momentum densities for the same wave packets. The color bar corresponds to the densities $|\psi(x,t)|^2$ (row 2) and $|\varphi(p,t)|^2$ (row 3). Row 2 insets – initial wave packet spatial densities.

The Airy and inverted Airy wave packets have identical position uncertainties, but different average arrival time delays, eliminating position uncertainty as the source of the differences in average arrival time delays. Therefore, either the momentum wave function phase or the spatial density profile must cause the difference in average arrival time delays. These two characteristics of the wave functions are inextricably linked, and their effects cannot be individually isolated. A change in momentum wave function phase always results in a change in spatial density and vice versa, and so we conclude that both the momentum wave function phase and spatial density profile control the interaction time.

Some insight into the role of these two features can be found by examining the phase time delay for the different wave packets, which is analogous to the average arrival time delay at the right edge of the barrier [13]. Equations (14) and (15) predict that the Airy (inverted Airy) will arrive after (before) the Gaussian wave packet with a difference in phase time delays of $\pm m \frac{(k_T - k_0)^2 - \alpha^2}{\hbar k_T}$. Using the mean transmitted momentum value from the numerical simulations ($k_T = 10.4$ a.u. for wave packets 1-3) yields a predicted difference in phase time delays of $\pm 0.7$ a.u. between the Gaussian and Airy/inverted Airy wave packets. These values are similar to the numerical results found in the simulation and shown in Table 1, confirming the relationship between average arrival time delay and phase time delay and demonstrating that the momentum wave function phase is the source of the time delay differences between the wave packets. If conditions are such that the non-Gaussian term in the phase time delay is negligible, then no difference in average arrival time delays will be observed, despite the differences in momentum wave function phase.

The clear influence of the momentum wave function phase on the average arrival time delay does not provide any information regarding the role of spatial density on the interaction time. To understand the effect of spatial density profile, it is necessary to examine the dwell times and self-interaction times of the different wave packets. Recall from Section 2.4 that the phase time delay can be written as the sum of the dwell time and self-interaction time. Therefore, the increased (decreased) interaction time with the barrier for the Airy (inverted Airy) wave packet relative to the Gaussian wave packet must be due to either or both of these components. Using Eq. (17), we calculated the dwell times for wave packets 1-3 and found them to be identical with $\tau_D = 0.03$ a.u, which is at least an order of magnitude smaller than the magnitude of the average arrival time delays. This makes the effect of the dwell time on the average arrival time delay negligible and implies that the average arrival time delay is largely determined by the self-interaction time, which therefore must differ by wave packet type. The self-interaction time is a result of interference between the incident and reflected parts of the wave packet, and insight into its variation with wave packet type can be found from the time-dependent momentum density plots in row 3 of Figure 4.

They show that the reflected part of the wave packet appears earliest for the inverted Airy wave packet, followed by the Gaussian and Airy wave packets. This is because the tail of the inverted Airy wave packet has significant non-zero density well to the right of the average position of the wave packet and it begins interacting with the barrier earlier than the leading edge of the Gaussian or Airy wave packets. Therefore, the reflection process begins first for the inverted Airy wave packet and last for the Airy wave packet. The momentum density plots also show that the small, positive momentum components persist for less time for the inverted Airy wave packet than for the Gaussian or Airy wave packets. This implies that for the inverted Airy wave packet, these components spend less time interacting with the barrier.

Additional information about the complementary roles of the spatial density and momentum wave function phase can be found through comparison of wave packets with very similar spatial densities, but different momentum wave function phases. Three such wave packets are listed as wave packets 4-6 in Table 1 and shown in Figure 5. These wave packets have identical spatial uncertainties, identical momentum densities, and nearly identical spatial densities, but different momentum wave function phases. Table 1 shows that the average arrival time delays for wave packets 4-6 are equal at $x' = 1$ a.u. and $x' = 20$ a.u. Calculation of their dwell times also results in identical values ($\tau_D = 0.02$ a.u.), and therefore, their self-interaction times must be identical. This confirms that while the momentum wave function phase is the source of the different interaction times, the wave packets must also have significantly different spatial densities in order to exhibit differences in interaction times.

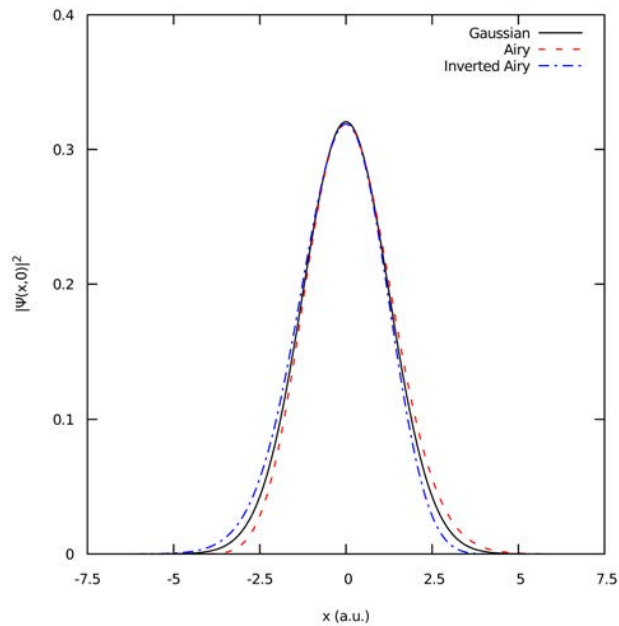

Figure 5 Spatial densities for wave packets 4-6 with $x_0 = 0$ a.u.

Lastly, for the wide barrier, we investigate the cause of the negative average arrival time delays. The acceleration of a wave packet due to its interaction with the barrier can be a result of two possible scattering mechanisms: tunneling or over-the-barrier scattering. It is well-known that tunneling wave packets arrive before their identical non-interacting counterparts [13,15,38,68], however in this case, the large barrier width suppresses tunneling. This is confirmed by the presence of only momentum components corresponding to energies above the barrier in the transmitted wave packet and is observable in the momentum density plots of Figure 4. Momentum values greater than 10 a.u. correspond to energies above the barrier height and are the primary components in the transmitted wave packets. Therefore, the negative average arrival time delays are due to the over-the-barrier scattering and not tunneling. The transmitted wave packets also each exhibit an increased average velocity (inverse slope, $\hbar k_T = 10.4$ a.u.) relative to the incident velocity (green line, $\hbar k_0 = 7.75$ a.u.), as seen by the change in slope of the transmitted wave packet trajectory in the spatial density plots. This is another indication that the large momentum components of the wave packet are the dominant contribution to transmission. The barrier effectively acts as a filter, allowing only the large momentum components to be transmitted.

### 3.2 Narrow Barrier

As shown above, the interaction times of the Airy, Gaussian, and inverted Airy wave packets differed due to their momentum wave function phases and spatial densities, but this was for one particular case when transmission was caused by over the barrier components of momentum and the contribution of tunneling to transmission was negligible. To determine if differences in interaction times exist for tunneling wave packets, we calculate time-dependent wave functions and interaction times for wave packets 1-3 with the same kinematics scattering from a narrow barrier with $w = 0.2$ a.u. Figure 6 shows time dependent spatial (row 2) and momentum (row 3) densities and arrival time distributions (row 1) for wave packets 1-3 scattering from the narrow barrier.

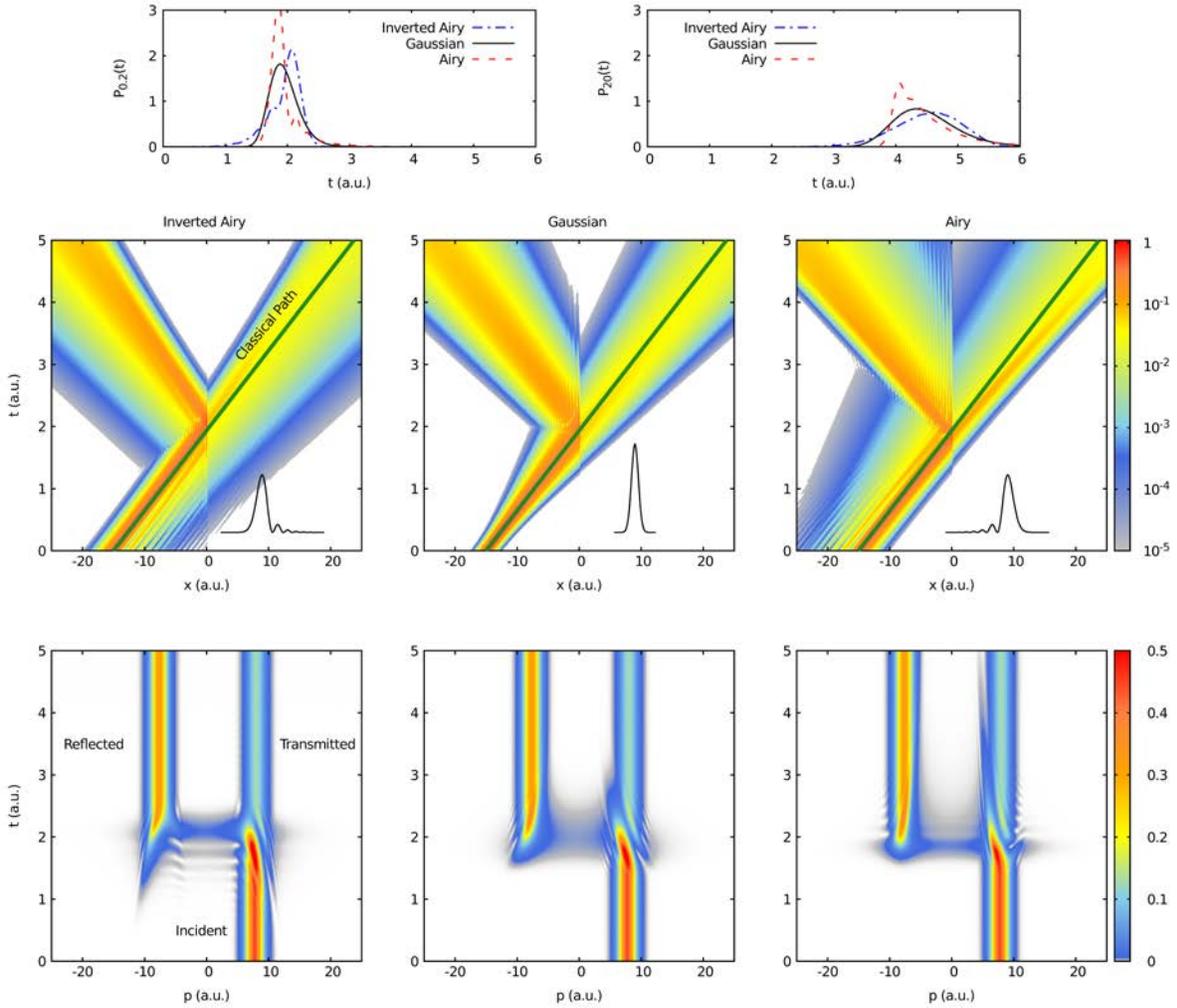

Figure 6 Same as Figure 4, but for a barrier width of $w = 0.2$ a.u.

The momentum densities show that for the tunneling wave packets, a larger range of momentum components are transmitted through the barrier with an average transmitted momentum of $\hbar k_T = 8$ a.u. for all three wave packets. While the tunneling wave packets' transmitted momenta are now more similar to the incident momentum, they are still larger, and therefore the tunneled wave packets arrive at $x' = 20$ a.u. before their free particle counterparts with negative average arrival time delays. The wider range of transmitted momentum values and the lower mean transmitted momentum are a result of more below-the-barrier momentum components tunneling through the barrier and contributing to the transmitted momentum. The similarity of transmitted and incident momentum values is also observed in the spatial densities of Figure 6 that show the slopes (inverse velocity) of the transmitted wave packet trajectories are similar to the slopes of the incident wave packets (green lines). The momentum density plots also show that as in the case of the wide barrier, the reflected part of the wave packet appears earliest for the inverted Airy wave packet, followed by the Gaussian and Airy wave packets. This is expected because this feature is caused by the spatial profile of the wave packets and should be independent of barrier width.

Unlike the arrival time distributions for the wide barrier, the arrival time distributions for the narrow barrier are overlapping and result in identical average arrival time delays that are close to zero ($\Delta T(x' = 0.2) = -0.01, \Delta T(x = -20) = -0.03$ a.u.). Calculation of the dwell times for the tunneling wave packets again yields identical values of $\tau_D = 0.02$ a.u. for each of the wave packets, which results in identical self-interaction times. At first glance, this seems contradictory to the results for scattering from the wide barrier, where the wave packets' spatial densities caused different self-interaction times and average arrival time delays. However, no contradiction exists because the spatial density profile only influences arrival times when the non-Gaussian term in the phase time delays of Eqs. (14) and (15) is significantly non-zero. In the case of tunneling wave packets, the similarity of the mean transmitted momentum and the mean incident momentum results in the non-Gaussian term in the phase time delays being negligible. Therefore, no difference in average arrival time delay is observed and consequently, there is no difference in self-interaction times. In this case, the spatial density of the wave packet is unimportant because the difference in phases does not lead to a difference in arrival times.

**3.3 Conditions for Average Arrival Time Delay Differences**

The results presented above were for wave packets with mean incident energy below the barrier height, and it is important to know how the interaction times change with incident energy. To this end, we calculate average arrival time delays, phase time delays, and dwell times as a function of incident momentum in three situations: (a) wave packets 1-3 scattering from the wide barrier, (b) wave packets 1-3 scattering from the narrow barrier, and (c) wave packets 4-6 scattering from the wide barrier.

In case (a), the wave packets have identical momentum densities but very different spatial densities and momentum wave function phases. Because the wide barrier is used, transmission is dominated by over-the-barrier momentum components and tunneling is negligible. Figure 7a and 7b show the average arrival time delays for case (a) as a function of incident energy. Generally, for wave packets with energies below the barrier height, the average arrival time delays are negative, while wave packets with energies above the barrier height have positive average arrival time delays. The one exception is the Airy wave packet average arrival time delay at the right edge of the barrier, which is positive for all incident energies. In this case, the Airy wave packet always arrives after its free particle counterpart, which is likely a result of the extended tail of the incident Airy wave packet to the left of its mean position.

The negative average arrival time delays for mean incident energy below threshold are expected because below-the-barrier scattering preferentially selects large momentum components from the incident wave packet resulting in the advance of an interacting wave packet relative to a free particle [13]. Wave packets with energies above the barrier will experience a reduced velocity in the barrier region, leading to positive average arrival time delays. The average arrival time delays of the wave packets are most different at low energy, with the Airy (inverted Airy) wave packet having the least (most) negative average arrival time delay. Further details on the differences in average arrival time delays can be observed in Figure 7c, which shows the relative average arrival time delays $(\Delta T^G - \Delta T^{A,IA})$ and relative phase time delays $(\Delta t_\zeta^G - \Delta t_\zeta^{A,IA})$ of the Airy and inverted Airy wave packets compared to the Gaussian wave packet. As incident energy increases, the mean transmitted momentum approaches the mean incident momentum, and the relative time delays go to zero, such that all three wave packets have identical interaction times. The dependence of the relative time delays on transmitted momentum is expected from the phase time delays of Eqs. (14) and (15) and is consistent with the special cases examined above. Only when the mean transmitted momentum differs from the mean incident momentum does the non-Gaussian term in the phase time delay result in non-zero relative average arrival time delays. The relative average arrival time delay curves in Figure 7c are approximately symmetric about zero, indicating that any advancement of the inverted Airy wave packet relative the Gaussian wave packet is nearly identical to the lag experienced by the Airy wave packet. Again, this is consistent with the predictions of Eqs. (14) and (15).

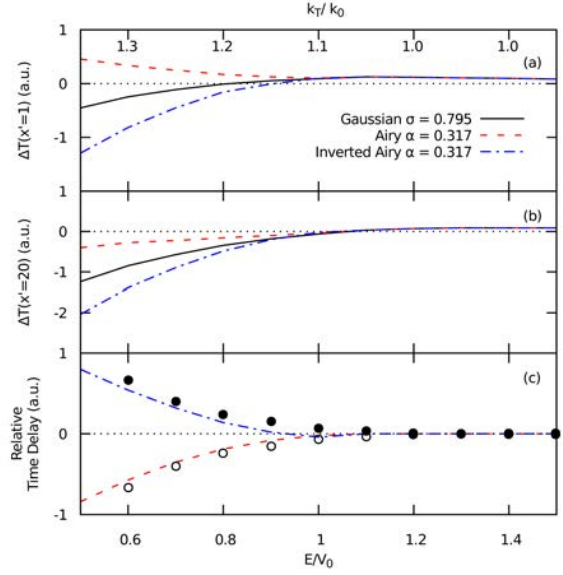

Figure 7 (a) Average arrival time delays for wave packets 1-3 at $x' = 1$ a.u. as a function of incident energy $E$ relative to barrier height $V_0$. The barrier width is $w = 1$ a.u. (b) Same as (a), but at $x' = 20$ a.u. (c) Relative average arrival time delays (lines) and relative phase time delays (points) of the Airy and inverted Airy wave packets as a function of incident energy. The top horizontal axis shows transmitted wave packet mean momentum $\hbar k_T$ relative to mean incident momentum $\hbar k_0$.

The dwell times for wave packets 1-3 are identical at each energy and shown as a function of incident energy in Figure 8. For case (a), the dwell time is a maximum for $\frac{E}{V_0} = 1.2$, where it is of similar magnitude to the average arrival time delay. Near threshold, the dwell time is the dominant contribution to the time delays and the self-interaction time is negligible. This is caused by reduced reflection probability for wave packets with larger incident energies, resulting in reduced interference between the incident and reflected parts of the wave packet. For wave packets with mean incident energies below the barrier height, the opposite is true. The dwell time is negligible and the self-interaction time is the dominant contribution to the time delays due to the increased probability of reflection that increases interference between the incident and reflected parts of the wave packet.

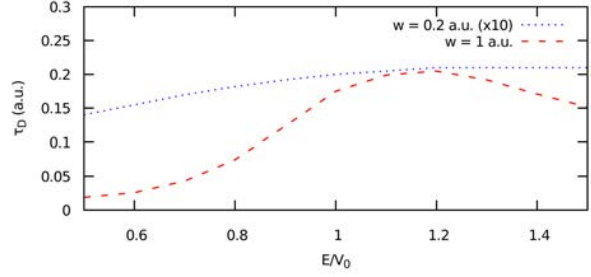

Figure 8 Dwell time as a function of incident energy for wave packets 1-3 interacting with barriers of width $w = 0.2$ and $w = 1$ a.u.

For case (b), we performed calculations of average arrival time delays, phase time delays, and dwell times for wave packets 1-3 scattering from the narrow barrier while varying the incident momentum ($7\ a.u. \leq \hbar k_0 \leq 13\ a.u.$). In these cases, tunneling is the dominant transmission mechanism and the average arrival time delays are close to zero ($|\Delta T(x' = 1)| <$ 0.02 a.u., $|\Delta T(x' = 20)| < 0.05$ a.u.). The ratio of the mean transmitted momentum to mean incident momentum is between 1 and 1.05 for all incident energies, which results in relative average arrival time delays and relative phase time delays also near zero. The dwell times for wave packets 1-3 were again identical at each incident energy and, as shown in Figure 8, exhibit a similar energy dependence as those of the wide barrier, except that they are an order of magnitude smaller. This again results in dwell time being the dominant contribution to average arrival time delays for wave packets with incident energies near threshold, but an insignificant contribution at low incident energies. Consequently, the self-interaction time is most important at low energies where more reflection occurs, however, it is independent of wave packet type due to the non-Gaussian term in the phase time delays being negligible.

For case (c), we performed calculations for wave packets 4-6 scattering from the wide barrier with incident momentum $7\ a.u. \leq \hbar k_0 \leq 13\ a.u.$ As in case (a), transmission here is largely due to over-the-barrier momentum components. However, unlike case (a), wave packets

4-6 have very similar spatial profiles, differing primarily only in momentum wave function phase. Based on the results above, we expect that the average arrival time delays will be similar in magnitude to those in case (a) and that if the ratio of mean transmitted momentum to mean incident momentum is approximately unity, the average arrival time delays will be identical for wave packets 4-6. This is exactly what is observed. The ratios of $k_T/k_0$ are between 1 and 1.09, leading to identical average arrival time delays at $x' = 20$ a.u. between -0.3 ($E/V_0 = 0.5$) and 0.25 ($E\backslash V_0 = 1.1$). The dwell times for wave packets 4-6 are also identical at each energy, closely resembling those in case (a). This again leads to the dwell time being the dominant contribution to time delays for wave packets with energies near threshold, while the self-interaction time is dominant for wave packets with energies below threshold.

Overall, cases (a) – (c) confirm that when the mean transmitted momentum is significantly different than the mean incident momentum, the non-linear terms in the momentum wave function phase lead to arrival times that are different from those of a Gaussian wave packet. The relative delay or advance of the non-Gaussian wave packet is caused by its altered self-interaction time that results from the wave packet's non-Gaussian spatial density profile. Differences in average arrival time delays are predominantly observed for wide barriers where transmission is dominated by over-the-barrier momentum components. For narrow barriers, where tunneling is the dominant transmission mechanism, no differences in average arrival time delays are observed.

The influence of the wave packet spatial profile and momentum wave function phase on the average arrival time delay provides a means for temporal control of wave packets with identical energies and momentum densities. This control is only possible with wave packets having non-linear momentum wave function phases, and in the next section, we demonstrate how such control may be achieved.

**3.4 Controlling Arrival Time**

The Airy and inverted Airy wave packets are two special cases of wave packets with cubic phases. To demonstrate the possibility of controlling average arrival time, we introduce a more general momentum wave function with a cubic phase containing an adjustable parameter $b$

$$\varphi^C(p,0) = \frac{\sqrt{\sigma}}{\pi^{1/4}} e^{-\sigma^2 (p-k_0)^2/2} e^{-ipx_0} e^{ib(p-k_0)^3}, \tag{18}$$

where $\sigma$ is the standard deviation of the Gaussian wave packet. The inverse Fourier transform of this wave function yields an initial state spatial wave function (see Figure 9) that can then be time evolved to study scattering dynamics and interaction times, as above. The stationary phase approximation predicts that the relative phase time delay between a Gaussian wave function and the wave function in Eq. (18) will be $-\frac{3b(k_T-k_0)^2}{\hbar k_T}$, which varies linearly with $b$. This implies that control of arrival time can be achieved by alteration of the cubic parameter $b$, independent of wave packet energy and momentum density.

In Figure 9 we show results from our numerical simulations and the stationary phase approximation for the relative average arrival time delay between the wave packet of Eq. (18) and a Gaussian wave packet as a function of $b$. Both wave packets have identical momentum densities with mean incident momentum $\hbar k_0 = 7.75$ a.u. and mean incident position $x_0 = -15$ a.u. The barrier width is $w = 1$ a.u. and barrier height is $V_0 = 50$ a.u. As predicted by the stationary phase approximation, Figure 9 shows that the relative average arrival time delay of the numerical calculations varies linearly with $b$ and confirms that the arrival time of a wave packet can be controlled through adjustment of this parameter. We note that the spatial wave function corresponding the momentum wave function of Eq. (18) resembles the Airy (positive $b$) or inverted Airy (negative $b$) wave packet with multiple peaks and larger spatial uncertainty than a Gaussian with identical momentum density. As $b$ increases, the spatial uncertainty increases, resulting in a larger self-interaction time that in turn leads to a larger relative average arrival time delay.

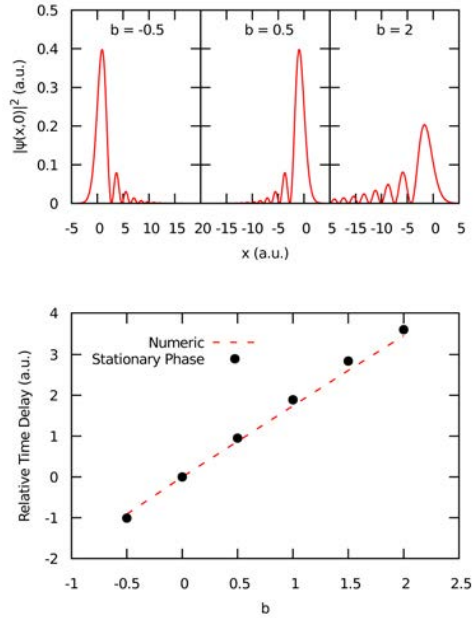

Figure 9 Top – initial wave packet spatial densities for the momentum wave function of Eq. (18). Bottom – relative average arrival time delays (red dashed line) and relative phase time delays (points) for the wave packet of Eq. (18). Mean incident energy is $\hbar k_0 = 7.75$ a.u., mean initial position is $x_0 = -15$ a.u., barrier width is $w = 1$ a.u., and barrier height is $V_0 = 50$ a.u.

Greater control over wave packet arrival time can be achieved through additional terms in the momentum wave function phase. For example, quadratic or linear terms could be included with independent parameters that provide further means of control. It is also possible to use polynomials higher than third order in the momentum wave function phase, which will yield more control parameters, each of which could be adjusted to provide the desired arrival time.

The influence of wave packet phase on average arrival time delay raises the question of whether the Hartman effect is present for non-Gaussian wave packets. For Gaussian wave packets, Hartman showed that the transmission time became independent of barrier width for wide barriers. By definition, the Hartman effect is based on the transmission time as defined by the phase time or group delay [38]. For non-Gaussian wave packets, we have shown that the phase time is simply the Gaussian phase time plus additional terms due to the non-linear parts the phase (see Eqs. (A14) and (A15)). These additional terms are independent of barrier width, leaving only the Gaussian term to vary with barrier width. Therefore, the Hartman effect is present for non-Gaussian wave packets (including the Airy and inverted Airy wave packets) and will have an identical functional dependence on barrier width, but a different asymptotic value of the saturated phase time due to the non-linear phase terms.

## 4. Conclusion

We have examined time-dependent scattering of non-Gaussian wave packets from one-dimensional square barrier potentials. We used average arrival time delay, phase time delay, and dwell time to quantify the wave packet's interaction time with the barrier. By comparing wave packets with identical momentum densities, we showed that the average arrival time delays of the transmitted wave packets are controlled by the momentum wave function phase. Under certain conditions, non-Gaussian wave packets exhibit different average arrival time delays than their Gaussian counterparts. When the mean transmitted momentum is significantly different than the mean incident momentum, the phase time delay of a non-Gaussian wave packet is shifted from that of a Gaussian wave packet due to the non-linear terms in its momentum wave function phase. Differences in the average arrival time delays are primarily observed for wave packets scattering from wide barriers, where transmission of over-the-barrier momentum components is the primary scattering mechanism. At low energies, the probability of reflection increases, which leads to an increased self-interaction time that becomes the dominant contribution to the time delay. The different spatial density profiles for Gaussian and non-Gaussian wave packets cause the self-interaction times to vary by wave packet type.

We have also demonstrated that the arrival time of a wave packet can be controlled by adjustment of the momentum wave function phase. Such control is only possible with non-Gaussian wave packets and numerical and analytical calculations for a wave function with an arbitrary cubic phase showed that the relative average arrival time delay was linearly dependent on the cubic term's coefficient, as predicted by the relative phase time delay. Thus, wave packets with identical incident energies and momentum densities can be designed to arrive on the far side of the barrier at differing and controllable times.

As interest in spatially structured electron beams grows, so too will the need to understand their dynamical properties. The difference in interaction times between the Gaussian and non-Gaussian wave packets shown here adds to the list of unique features exhibited by these wave packets and may open the door to new applications. Current experimental techniques allow for measurements on the timescales of a few attoseconds, making the relative average arrival time delays predicted here large enough to be measured. Thus, an experiment comparing average arrival time delays of Airy and Gaussian wave packets could provide insight into the validity of the various arrival time definitions.

**Acknowledgements**
We gratefully acknowledge the support of the NSF under Grant No. PHY-1912093.

## Appendix A

We are interested in calculating the phase time delays of Airy, inverted Airy, and Gaussian wave packets using the stationary phase approximation. Following the derivation of [20], a time-evolved wave packet initially centered at $x = x_0$ may be written as

$$\psi(x,t) = \frac{1}{\sqrt{2\pi}} \int_{-\infty}^{\infty} \phi(p,0)\, u_p(x) e^{-i\frac{\hbar p^2}{2m}t} dp \tag{A1}$$

where $\phi(p,0)$ is the initial momentum wave function and $u_p(x)$ is an energy eigenstate with momentum $p$. For the transmitted wave packet to the right of the barrier, the energy eigenstate can be written as

$$u_p(x) = C e^{ipx}, \tag{A2}$$

where $C = -\frac{e^{-\rho L} 2i \sin 2\theta}{R} e^{-ipL} e^{-i\beta}$ and $|C|^2$ is the transmission coefficient. The quantities $\rho, R, \theta$, and $\beta$ are expressed in terms of $p$ and incident wave number $k_0$

$$\rho^2 = k_0^2 - p^2 \tag{A3}$$

$$\cos\theta = \frac{p}{k_0} \tag{A4}$$

$$\sin\theta = \frac{\rho}{k_0} \tag{A5}$$

$$R e^{-i\beta} = e^{2i\theta} e^{-2\rho L} - e^{-2i\theta} \tag{A6}$$

$$R = \sqrt{1 + e^{-4\rho L} - 2e^{-2\rho L} \cos 4\theta} \tag{A7}$$

$$\tan\beta = \coth\rho L \tan 2\theta. \tag{A8}$$

Inserting the momentum wave function for the Gaussian wave packet of Eq. (4) into Eq. (A1) yields

$$\psi^G(x,t) = \frac{-1}{\sqrt{2\pi}} \int_{-\infty}^{\infty} \frac{\sigma^{\frac{1}{2}}}{\pi^{1/4}} e^{-ipx_0} e^{-\sigma^2(p-k_0)^2/2} \frac{e^{-2\rho L} 2i \sin 2\theta}{R} e^{-ipL} e^{-i\beta} e^{ipx} e^{-i\frac{\hbar p^2}{2m}t} dp. \tag{A9}$$

By assuming that the phase $f$ of the wave packet in Eq. (A9) is stationary at the right edge of the barrier ($x = L$), an equation of motion can be found that yields a time of arrival at the right edge of the barrier, i.e.

$$\left.\frac{df}{dp}\right|_{k_T} = 0, \tag{A10}$$

where $\hbar k_T$ is the mean momentum of the transmitted wave packet. For the Gaussian wave packet, the phase is

$$f = p(x - x_0 - L) - \frac{\pi}{2} - \beta - \frac{\hbar p^2 t}{2m} \tag{A11}$$

and the corresponding equation of motion is

$$x = x_0 + L + \left.\frac{d\beta}{dp}\right|_{k_T} + \frac{\hbar k_T t}{m}. \tag{A12}$$

Setting $x = L$ and solving for $t$ gives

$$t_L^G = \frac{-m\left(x_0 + \left.\frac{d\beta}{dp}\right|_{k_T}\right)}{\hbar k_T}. \tag{A13}$$

Similar derivations for the Airy and inverted Airy wave packets yield

$$t_L^A = \frac{m\left[-x_0 - \left.\frac{d\beta}{dp}\right|_{k_T} + (k_T - k_0)^2 - \alpha^2\right]}{\hbar k_T} = t_L^G + \frac{m[(k_T - k_0)^2 - \alpha^2]}{\hbar k_T} \tag{A14}$$

and

$$t_L^{IA} = \frac{m\left[-x_0 - \left.\frac{d\beta}{dp}\right|_{k_T} - (k_T - k_0)^2 + \alpha^2\right]}{\hbar k_T} = t_L^G - \frac{m[(k_T - k_0)^2 - \alpha^2]}{\hbar k_T}. \tag{A15}$$

These are the stationary phase times for wave packet arrival at the right edge of the barrier. The corresponding phase time delays can be found by subtracting the non-interacting wave packet arrival time at $x = L$ from the scattered wave packet arrival time

$$\Delta t_\zeta^i = t_L^i - \frac{m(L - x_0)}{\hbar k_0}, \tag{A16}$$

where $i = G, A, IA$ for Gaussian, Airy, and inverted Airy wave packets. Using Eqs. (A13) - (A16), the phase time delays can be written as in Eqs. (14) and (15)

$$\Delta t_\zeta^A = \Delta t_\zeta^G + m\frac{(k_T - k_0)^2 - \alpha^2}{\hbar k_T} \tag{A17}$$

$$\Delta t_\zeta^{IA} = \Delta t_\zeta^G - m\frac{(k_T - k_0)^2 - \alpha^2}{\hbar k_T}. \tag{A18}$$